\documentclass[preprint,12pt]{elsarticle}
\usepackage{amsmath,amssymb,amsfonts}
\usepackage{algorithmic}
\usepackage{adjustbox}
\usepackage{algorithm}
\usepackage{empheq}
\usepackage{graphicx}
\usepackage{multirow}
\usepackage{mfirstuc}
\usepackage{textcomp}
\usepackage{xcolor}
\usepackage{amssymb}
\usepackage{svg}
\usepackage{amsmath}
\usepackage{caption}
\usepackage{subcaption}
\usepackage{amsfonts}
\usepackage{url}
\usepackage{color}
\usepackage{ragged2e}
\definecolor{blue}{rgb}{0,0,1}

\mathchardef\mhyphen="2D
\usepackage{soul}
\usepackage{lineno,hyperref}

\let\OLDthebibliography\thebibliography
\renewcommand\thebibliography[1]{
	\OLDthebibliography{#1}
	\setlength{\parskip}{0pt}
	\setlength{\itemsep}{0pt plus 0.3ex}
}
\begin{document}
	
	\captionsetup[figure]{name=Fig. ,labelsep=period}
	
	\begin{frontmatter}
		
		\title{Multi-Objective Planning of Community Energy Storage Systems Under Uncertainty}
		
		\author[1]{K.B.J. Anuradha\corref{cor1}}
		\ead{Jayaminda.KariyawasamBovithanthri@anu.edu.au}
		\author[1]{José Iria}
		\ead{Jose.Iria@anu.edu.au}
		\author[1,2]{Chathurika P. Mediwaththe}
		\ead{chathurika.mediwaththe@csiro.au}
		
		\address[1]{College of Engineering and Computer Science, The Australian National University (ANU), Canberra, ACT 2601, Australia.}
		\address[2]{Commonwealth Scientific and Industrial Research Organisation (CSIRO), Canberra, ACT 2601, Australia.}
		\cortext[cor1]{Corresponding author.}

		\makeatletter
		\def\ps@pprintTitle{%
		\let\@oddhead\@empty
		\let\@evenhead\@empty
		\let\@oddfoot\@empty
		\let\@evenfoot\@oddfoot
		}
		\makeatother
		
		\begin{abstract}
			This paper evaluates how the planning of a community energy storage (CES) system under different energy trading schemes (ETSs) can benefit low voltage (LV) prosumers and the CES provider equitably. First, we consider an ETS where the CES provider trades energy with prosumers at the average grid energy trading price, second, an ETS where the CES provider trades energy at a higher price than the grid energy trading price, and third, an ETS where the CES provider trades energy at a lower price than the grid energy trading price. To this end, we present a multi-objective stochastic optimization framework to minimize the investment and annual operating costs of the CES provider and annual operating costs of prosumers, taking into account the uncertainties of real and reactive energy consumption and photovoltaic (PV) generation of prosumers. The uncertainties are modeled using the normal probability density function. Then, the roulette wheel mechanism (RWM) is exploited to formulate a scenario-based stochastic program. The initial scenarios obtained from the RWM, are then reduced using the K-Means clustering algorithm, to make the problem tractable. Our experiments show that the ETS where the CES provider trades energy at the average grid energy trading price benefits prosumers and the CES provider more equitably than the other two ETSs.
			
		\end{abstract}
		
		
	\end{frontmatter}
	
	\section{{\normalsize Introduction}}
	
	\subsection{{\normalsize Motivation}}
	
	The integration of photovoltaic (PV) systems in low voltage (LV) grids has significantly increased over the last few years. However, the intermittent and non-dispatchable nature of PV generation may restrict exploiting its merits fully. Community energy storage (CES) systems are an emerging type of battery system, which can efficiently accommodate those features of PV generation. Typically, a CES system trades energy with multiple prosumers, and the grid to economically and/or technically benefit them. Also, the CES technology is gaining increasing interest in the industry and research community, as it can enable increased community access and network hosting capacity for renewable energy \cite{shaw2020community}.
	
	An energy management framework that only aims at optimizing the CES system operation, may not deliver the expected rewards from it completely. Hence, it is imperative that the planning aspects including the location, the capacity, and the rated power of a CES system are optimized concurrently with its operation. Additionally, energy management problems involving a CES system may have competing objectives among different stakeholders such as minimizing the investment and operating costs of the CES system for the CES provider and minimizing the operating costs for prosumers. Also, it is essential to value the energy trade by a CES system to economically benefit its provider and/or prosumers suitably. Thus, a multi-objective optimization framework with a properly selected CES provider's energy trading scheme (ETS) can help decision makers such as utilities, energy regulators, and community leaders to identify the trade-offs between those objectives, and thereby benefit both prosumers and the CES provider.  
	\vspace{-0.5em}
	\subsection{{\normalsize Related work}}
	
	Several studies have presented methods to optimize the CES system operation \cite{mediwaththe2020network,mediwaththe2020community}, and both the CES system planning and operation \cite{divshali2017improving}-\cite{bohringer2021sizing}. For instance, authors of \cite{mediwaththe2020network} have presented a framework to optimize the CES system operation, to minimize prosumers' operating costs while maximizing the revenue of the CES provider. In \cite{mediwaththe2020community}, the operation of a CES system is optimized to minimize the real energy losses and energy trading costs with the grid by the CES provider and prosumers. The CES system planning and operation are optimized simultaneously to enhance the hosting capacity in \cite{divshali2017improving}, and to mitigate voltage excursions in unbalanced LV grids in  \cite{alam2015community}. Analytical methods based on graphical and sensitivity analyses for optimizing the CES system planning and operation have been discussed in \cite{hung2011community,bohringer2021sizing}. A common feature of \cite{mediwaththe2020network}-\cite{bohringer2021sizing} is that authors have used deterministic models, assuming the PV generation and real energy consumption of prosumers are not subject to uncertainty. Nevertheless, neglecting the uncertainty of PV generation and real energy consumption may result in models presented in \cite{mediwaththe2020network}-\cite{bohringer2021sizing} not being efficient in providing cost-effective planning and operation decisions for the CES system, as the forecast errors can be quite high.
	
	In \cite{Liu2023robust}-\cite{SalazarDuque2022community}, the uncertainty of real energy consumption and PV generation has been taken into account for energy management problems involving a CES system. A game-theory approach for optimizing the energy management of a CES system considering the uncertainty of PV and wind power generation is discussed in \cite{Liu2023robust}. Authors of \cite{pamshetti2020coordinated} have presented a method to optimize the CES system planning to accommodate the high PV generation of LV grids and mitigate voltage excursions. The authors have used a normal distribution to model uncertainties of PV generation and real energy consumption, and then the roulette wheel mechanism to generate scenarios. In \cite{mahmoodi2020voltage}, authors have investigated how the CES system location impacts the voltage profile and energy losses in LV networks. Moreover, methods to optimize the CES system operation under the uncertainties of PV generation and real energy consumption of prosumers are discussed in \cite{Parvar2023optimal, SalazarDuque2022community}.
	
	In summary, the research works in \cite{mediwaththe2020network}-\cite{SalazarDuque2022community} have not explored how different ETSs impact prosumers' and the CES provider's economic benefits gained by optimizing the CES system planning and operation under a multi-objective stochastic framework. In contrast to the existing literature, we optimize the CES system planning and operation under a multi-objective stochastic framework to select the ETS that economically benefits prosumers and the CES provider most equitably. By saying equitably, we mean a balanced distribution of economic benefits between prosumers and the CES provider.
	
	\vspace{-0.5em}
	\subsection{{\normalsize Contributions}}
	
	In this paper, we examine the extent to which the planning and operation of a CES system can benefit prosumers and the CES provider equitably. To this end, we present a multi-objective stochastic optimization framework to size and place a CES system under three ETSs with different energy trading prices for the CES provider. This multi-objective stochastic framework minimizes the investment and operating costs of the CES provider and the operating costs of prosumers under three ETSs, helping decision-makers to select the solution that delivers the most equitable economic benefits to prosumers and the CES provider. We model the uncertainties of PV generation, and real and reactive energy consumption using stochastic scenarios generated initially by a roulette wheel mechanism (RWM) and then aggregated by the K-Means clustering algorithm to make the optimization problems tractable.
	
	In summary, the main contributions of this paper are the following:
	
	\vspace{-0.5em}
	\begin{itemize}
		\item We present a multi-objective stochastic optimization framework to size and place a CES system for three ETSs with different energy trading prices for the CES provider. The aim of the multi-objective stochastic optimization framework is to help decision-makers, including energy regulators, utilities, and community leaders to select the solution that delivers the most economically equitable outcome to the CES provider and prosumers. To the best of our knowledge, there is no study in the literature that proposes to use a multi-objective stochastic optimization framework to help stakeholders select a CES solution (optimal CES system planning and the ETS) that delivers economically equitable outcomes to prosumers and the CES provider.
		\vspace{-0.5em}
		\item Our multi-objective stochastic optimization framework allows evaluating multi-objective impacts of uncertainties of PV generation, and real and reactive energy consumption in the planning of CES systems. This contrasts with the stochastic approaches \cite{Liu2023robust}-\cite{SalazarDuque2022community} which do not evaluate the multi-objective impacts of all mentioned uncertainty parameters, and with the deterministic approaches discussed in \cite{mediwaththe2020network}-\cite{bohringer2021sizing} which do not evaluate any uncertainty impact.
		\vspace{-0.5em}
		\item Our experiments on an LV distribution network with 30 real-world prosumers from an energy community prove that the proposed multi-objective stochastic optimization framework benefits the CES provider most when the CES provider's energy trading price is higher than the grid energy price, and prosumers are benefited most when the CES provider's energy trading price is lower than the grid energy trading price. Also, among the three ETSs, the most equitable economic benefits for prosumers and the CES provider are attained when the CES provider trades energy at the average grid energy trading price.
		
	\end{itemize}
	\vspace{-0.5em}
	\subsection{{\normalsize Paper organization}}
	
	This paper is structured as follows. The system models of our paper are presented in Section 2. The stochastic modeling is described in Section 3. The formulation of the multi-objective stochastic optimization framework is described in Section 4. Section 5 presents the numerical and graphical results, and Section 6 gives the conclusion of the paper.
	\vspace{-1mm}
	\section{{\normalsize System modeling}}
	
\begin{figure}[b]
		\centering
		\includegraphics[width=3.5in,height=1.8in]{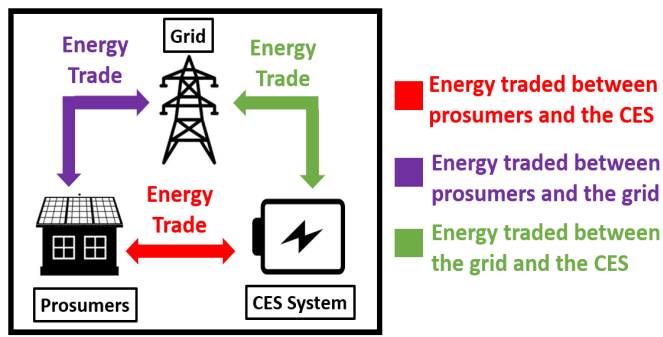}
		\caption{Energy trade between prosumers, the grid, and the CES system.}\label{fig: Fig. 1}
		\vspace{-1em}
	\end{figure}

	In this paper, prosumers, the CES system, and the grid can trade energy between them, as illustrated in Fig. \ref{fig: Fig. 1}. This section details models of prosumers, network power flow, energy trading schemes, and the CES system.
	\vspace{-0.5em}
	\subsection{{\normalsize Preliminaries}}
	
	Prosumers $\mathcal{U}$ are assumed to own PV units. We assume the CES system is owned by a third party, and we designate its owner as the CES provider. Also, all the analyses in this paper are done at time $t \in \mathcal{T} $, where $\mathcal{T}$ is the set of time instances, and $\Delta t$ is the difference between two consecutive time instances. Moreover, we consider the uncertainty of PV generation and the energy consumption (real and reactive) of prosumers. To model the uncertainty, we use three normal probability density functions (PDFs) each for real and reactive energy consumption and PV generation of prosumers, which are then discretized to formulate a scenario-based stochastic program. Due to the computational complexity of such a stochastic program, it is imperative to use a scenario reduction approach to reduce the number of scenarios to keep the problem tractable. For this, we use the K-Means clustering algorithm. In this work, the initial set of scenarios is denoted by $\mathcal{M}$, and the set of scenarios obtained after applying the K-Means algorithm is given by $ \mathcal{S}$. Further explanation of the uncertainty modeling is given in Section 3.
	\vspace{-2mm}
	\subsection{{\normalsize Prosumers' model}}
	
	As seen in Fig. \ref{fig: Fig. 1}, $ \forall u \in \mathcal{U}, \hspace{1mm}  t \mathcal{\in T}, \hspace{1mm} s \mathcal{\in S}$, prosumers trade energy $e_{u,s}^{G}(t)$ with the grid, and $e_{u,s}^{CES}(t)$ the CES system to balance the mismatch of their PV energy generation $e_{u,s}^{PV}(t)$ and real energy consumption $e_{u,s}^{L}(t)$. Prosumers import energy from the grid and/or the CES system when $e_{u,s}^{L}(t) > e_{u,s}^{PV}(t)$ as described by (\ref{eq:1}), and export the energy back to them when $e_{u,s}^{L}(t) < e_{u,s}^{PV}(t)$ as given in (\ref{eq:2}). For prosumers' energy imports from the grid, $e_{u,s}^{G}(t) > 0$, and $e_{u,s}^{G}(t) < 0$ for their energy exports to the grid. The same sign convention is used for energy trades with the CES system by prosumers.
	\vspace{-0.5em}
	\begin{subequations}\label{eq:1}
		\begin{equation}\label{eq:1a}
			0 \leq e_{u,s}^{G}(t)+ e_{u,s}^{CES}(t) = e_{u,s}^{L}(t)-e_{u,s}^{PV}(t) 
		\end{equation}
		\vspace{-0.5em}
		\begin{equation}\label{eq:1b}
			\begin{split}
				0 \leq e_{u,s}^{G}(t) \leq  e_{u,s}^{L}(t)-e_{u,s}^{PV}(t) \hspace{5mm} \forall  u \in \mathcal{U}, \hspace{1mm}  t \mathcal{\in T}, \hspace{1mm} s \mathcal{\in S}	\hspace{3mm} if \hspace{3mm} e_{u,s}^{L}(t) > e_{u,s}^{PV}(t)
			\end{split}
		\end{equation}
	\end{subequations}
	\vspace{-0.5em}
	\begin{subequations}\label{eq:2}
		\begin{equation}\label{eq:2a}
			e_{u,s}^{G}(t)+ e_{u,s}^{CES}(t) = e_{u,s}^{L}(t)-e_{u,s}^{PV}(t) \leq 0 \hspace{10mm}  
		\end{equation}
		\vspace{0.5mm}
		\begin{equation}\label{eq:2b}
			\begin{split}
				e_{u,s}^{L}(t)-e_{u,s}^{PV}(t) \leq e_{u,s}^{G}(t) \leq 0 \hspace{5mm} \forall  u \in \mathcal{U}, \hspace{1mm}  t \mathcal{\in T}, \hspace{1mm} s \mathcal{\in S}	\hspace{3mm} if \hspace{3mm} e_{u,s}^{L}(t) < e_{u,s}^{PV}(t)
			\end{split}
		\end{equation}
	\end{subequations}
	
	\subsection{{\normalsize Network power flow model}}
	
	In this paper, a distribution network with a radial topology is considered. It is described by the graph $\mathcal{G=(V,E)}$, where $\mathcal V =\left \{{0,1,..., N} \right \}$  is the set of all nodes, and $\mathcal E= \left \{ (\alpha,\beta) \right \}\subset \mathcal V\times \mathcal V $ is the set of all lines in the  network. Node 0 (slack node) represents the secondary side of the distribution transformer. The resistance and the reactance of line $(\alpha,\beta)$ are $r_{\alpha,\beta}$ and $x_{\alpha,\beta}$, respectively. For scenario $s$ at time $t$, real and reactive power flows from $\alpha$ to $\beta$ node are represented by $P_{\alpha,\beta,s}(t)$ and $Q_{\alpha,\beta,s}(t)$, respectively. Furthermore, $\forall \alpha\in \mathcal{V\setminus}\left \{ 0 \right \}$, $t \mathcal{\in T}$, $ s \mathcal{\in S}$, real power absorption, reactive power absorption, voltage magnitude are given by $p_{\alpha,s}(t)$, $q_{\alpha,s}(t)$, and $V_{\alpha,s}(t)$, respectively. The linear Distflow equations (\ref{eq:3})-(\ref{eq:5}) are used to model power flows in the radial network \cite{Baran1989optimal}. These linear Distflow equations compute good results in small LV networks since the relaxed losses are small. Let $U_{\alpha,s }(t)= V^2_{\alpha,s}(t) $.
	\vspace{-0.5em}
	\begin{equation}\label{eq:3}
		\begin{split}    
			P_{\alpha,\beta,s}(t)=p_{\alpha,s}(t)+\sum_{\gamma:\beta\rightarrow \gamma}^{}P_{\beta,\gamma,s}(t) \hspace{4mm} \forall (\alpha,\beta) \in \mathcal{E}, \hspace{1mm} t \mathcal{\in T} \hspace{1mm}, \hspace{1mm} s \mathcal{\in S}
		\end{split}
	\end{equation}
	\vspace{-1em}
	\begin{equation}\label{eq:4}
		\begin{split}
			Q_{\alpha,\beta,s}(t)=q_{\alpha,s}(t)+\sum_{\gamma:\beta\rightarrow \gamma}^{}Q_{\beta,\gamma,s}(t) \hspace{4mm} \forall (\alpha,\beta) \in \mathcal{E}, \hspace{1mm} t \mathcal{\in T} \hspace{1mm}, \hspace{1mm} s \mathcal{\in S}
		\end{split}
	\end{equation}
	\vspace{-0.5em}
	\begin{equation}\label{eq:5}
		\begin{split}
			U_{\beta,s }(t)=U_{\alpha,s }(t)-2(r_{\alpha,\beta}P_{\alpha,\beta,s}(t)+x_{\alpha,\beta}Q_{\alpha,\beta,s}(t)) \hspace{3mm} \forall (\alpha,\beta) \in \mathcal{E}, \hspace{1mm} t \mathcal{\in T} \hspace{1mm}, \hspace{1mm} s \mathcal{\in S}
		\end{split}
	\end{equation}
	
	The real and reactive power absorbed by the node $\alpha$ at time $t$ in scenario $s$ can be expressed by (\ref{eq:6}) and (\ref{eq:7}). Here, we assume there are multiple prosumers at each node, and the CES system operates at a unity power factor. We denote prosumers at node $\alpha$ by $\mathcal{U}_{\alpha}  \subset \mathcal{U}$. The actual energy flowing in and out of the CES system is represented by $e_{\alpha,s}^{CES,ch}(t)$ for charging and $e_{\alpha,s}^{CES,dis}(t)$ for discharging. Let  $e_{u,s}^{Q}(t)$ be the reactive energy consumption per prosumer.
	\vspace{-0.5em}
	\begin{equation}\label{eq:6}
		\begin{split}
			p_{\alpha,s}(t) =\frac{1}{\Delta t}(\sum_{u \in \mathcal{U_{\alpha}}}^{}e_{u,s}^{L}(t)-\sum_{u \in \mathcal{U_{\alpha}}}^{}e_{u,s}^{PV}(t)+ e_{\alpha,s}^{CES,ch}(t) - e_{\alpha,s}^{CES,ch}(t)) \\ \forall \alpha \in \mathcal{V\setminus}\left \{ 0 \right \}, \hspace{1mm} t \mathcal{\in T}, 	\hspace{1mm} s \mathcal{\in S}	\hspace{7mm}
		\end{split}		
	\end{equation}
	\vspace{-0.5em}
	\begin{equation}\label{eq:7}
		\begin{split}
			q_{\alpha,s}(t) =\frac{1}{\Delta t}(\sum_{u \in \mathcal{U_{\alpha}}}^{}e_{u,s}^{Q}(t)) \hspace{3mm} \forall \alpha \in \mathcal{V\setminus}\left \{ 0 \right \}, \hspace{1mm} t \mathcal{\in T}, \hspace{1mm} s \mathcal{\in S}	
		\end{split}		
	\end{equation}
	
	We use (\ref{eq:8}) to guarantee the voltage magnitude of nodes are within lower ($ \underline{V} $) and upper  ($ \overline{V} $) bounds, where $\underline{U}=\ \underline{V}^2 $ and $\overline{U}= \overline{V}^2 $. The line thermal limits are not explicitly modeled, as they do not constrain the optimization problems in our case study. However, they can be easily added if required.
	\vspace{-0.5em}
	\begin{equation}\label{eq:8}
		\underline{U} \le U_{\alpha,s }(t) \le \overline{U} \hspace{8mm} \forall \alpha \in \mathcal{V\setminus}\left \{ 0 \right \}, \hspace{1mm} t \mathcal{\in T}, \hspace{1mm} s \mathcal{\in S}	
	\end{equation}
	\vspace{-10mm}
	\subsection{{\normalsize Community energy storage system model}}
	
	The CES system planning and operational aspects are modeled by (\ref{eq:9})-(\ref{eq:11}) and (\ref{eq:12})-(\ref{eq:15}), respectively. The optimal CES system at node $\alpha$ is given by capacity $E_{\alpha}^{CES,C}$ and rated power $p_{\alpha}^{CES,R}$. The binary variable $A_{\alpha}$ is 1 when $\alpha$ is the optimal CES system node. The minimum and maximum CES capacity limits and the maximum CES power are denoted by $\underline{E}^{CES}$, $\overline{E}^{CES}$ and $\overline{p}^{CES}$, respectively. The charging and discharging of the CES system are modeled by (\ref{eq:12a})-(\ref{eq:12f}). We use the binary variable $B_{\alpha,s}(t)$ and two continuous variables $m_{s}(t)$ $\in \mathbb{R}^{+}$ and $n_{s}(t)$ $\in \mathbb{R}^{+}$ to make sure charging and discharging of the CES system do not occur simultaneously. When the CES system charges, $B_{\alpha,s}(t)=1$, and thus, $0 \leq e_{\alpha,s}^{CES,ch}(t) \leq m_{s}(t)=p_{\alpha}^{CES,R}\Delta t \leq \overline{p}^{CES}\Delta t$. When the CES system discharges, $B_{\alpha,s}(t)=0$, and hence, $0 \leq e_{\alpha,s}^{CES,dis}(t) \leq n_{s}(t)=p_{\alpha}^{CES,R}\Delta t \leq \overline{p}^{CES}\Delta t$. The variation of the CES system energy level with time is modeled by (\ref{eq:13}), the CES system energy level is maintained within the minimum and maximum allowable energy levels by (\ref{eq:14}), and the CES system continuous operation over the next day is facilitated by (\ref{eq:15}). In (\ref{eq:13}) and (\ref{eq:14}), $E_{\alpha,s}^{CES}(t)$ is the CES system energy level at time $t$ in scenario $s$. The parameters $\mu ^{ch},\mu^{dis}$ and $\underline{\sigma }, \overline{\sigma }$ are the charging and discharging efficiencies of the CES system, and the minimum and maximum percentage coefficients of the CES system capacity, respectively. Also, $\theta $ is a small positive number, and $t_{p} \in \mathcal{T}_{P}$ where $\mathcal{T}_{P}=\left \{ 1,2,....,\left |  \mathcal T \right |/24\right \}$. 
	\vspace{-0.5em}
	\begin{equation}\label{eq:9}
		\sum_{\alpha=1}^{N}A_{\alpha}=1  \hspace{5mm} \forall \alpha \in \mathcal{V\setminus}\left \{ 0 \right \},  A_{\alpha} \in \left \{ 0,1 \right \} 
	\end{equation}
	\vspace{-0.5em}
	\begin{equation}\label{eq:10}
		A_{\alpha} \underline{E}^{CES}  \le  E_{\alpha}^{CES,C} \le A_{\alpha} \overline{E}^{CES}  \hspace{5mm} \forall \alpha \in \mathcal{V\setminus}\left \{ 0 \right \},  A_{\alpha} \in \left \{ 0,1 \right \} 
	\end{equation}	
	\vspace{-0.7em}	
	\begin{equation}\label{eq:11}
		0 \le  p_{\alpha}^{CES,R} \le A_{\alpha} \overline{p}^{CES}    \hspace{5mm} \forall \alpha \in \mathcal{V\setminus}\left \{ 0 \right \},  A_{\alpha} \in \left \{ 0,1 \right \}
	\end{equation}
	\vspace{-0.8em}
	\begin{subequations}\label{eq:12}
		\begin{equation}\label{eq:12a}
			0 \leq e_{\alpha,s}^{CES,ch}(t) \leq m_{s}(t)
		\end{equation}
		\vspace{-0.8em}
		\begin{equation}\label{eq:12b}
			e_{\alpha,s}^{CES,ch}(t) \leq m_{s}(t) \leq \overline{p}^{CES}\Delta t B_{\alpha,s}(t)
		\end{equation}
		\vspace{-0.8em}
		\begin{equation}\label{eq:12c}
			-\overline{p}^{CES}\Delta t (1-B_{\alpha,s}(t)) \leq m_{s}(t)-p_{\alpha}^{CES,R}\Delta t\leq 0 
		\end{equation}
		\vspace{-0.8em}
		\begin{equation}\label{eq:12d}
			0 \leq  e_{\alpha,s}^{CES,dis}(t) \leq n_{s}(t)
		\end{equation}
		\vspace{-0.7em}
		\begin{equation}\label{eq:12e}
			e_{\alpha,s}^{CES,dis}(t) \leq n_{s}(t) \leq \overline{p}^{CES}\Delta t (1-B_{\alpha,s}(t))
		\end{equation}
		\vspace{-0.5em}
		\begin{equation}\label{eq:12f}
			\begin{split}
				-\overline{p}^{CES}\Delta t B_{\alpha,s}(t) \leq n_{s}(t)-p_{\alpha}^{CES,R}\Delta t \leq 0 \hspace{6mm} \forall \alpha \in \mathcal{V\setminus}\left \{ 0 \right \}, \hspace{1mm} t \mathcal{\in T}, \hspace{1mm} s \mathcal{\in S} \hspace{10mm}
			\end{split}	 
		\end{equation}
	\end{subequations}
	\vspace{-0.5em}
	\begin{equation}\label{eq:13}
		\begin{split}
			E_{\alpha,s}^{CES}(t) = E_{\alpha,s}^{CES}(t-1)+\mu^{ch}e_{\alpha,s}^{CES,ch}(t) -\frac{1}{\mu^{dis}}e_{\alpha,s}^{CES,dis}(t) \\ \forall \alpha \in \mathcal{V\setminus}\left \{ 0 \right \}, \hspace{1mm} t \mathcal{\in T}, \hspace{1mm} s \mathcal{\in S} \hspace{10mm}
		\end{split}
	\end{equation}
	\vspace{-0.5em}
	\begin{equation}\label{eq:14}
		\begin{split}
			\underline{\sigma } E_{\alpha}^{CES,C} \le E_{\alpha,s}^{CES}(t) \le \overline{\sigma } E_{\alpha}^{CES,C} \hspace{4mm} \forall \alpha \in \mathcal{V\setminus}\left \{ 0 \right \}, \hspace{1mm} t \mathcal{\in T}, \hspace{1mm} s \mathcal{\in S}
		\end{split}
	\end{equation}
	\vspace{-5mm}
	\begin{equation}\label{eq:15}
		\begin{split}
			\left |  E_{\alpha,s}^{CES}(24 t_{p})-E_{\alpha,s}^{CES}(0) \right |\leq \theta \hspace{4mm} \forall \alpha \in \mathcal{V\setminus}\left \{ 0 \right \}, \hspace{1mm} t_{p} \mathcal{\in T}_{P}, \hspace{1mm} s \mathcal{\in S}
		\end{split}		
	\end{equation}
	\vspace{-5mm}
	\subsection{{\normalsize Energy trading schemes}}
	
	The CES provider trades energy with prosumers at a price $\lambda _{C}(t)$, and the grid trades energy with prosumers and the CES system at a price $\lambda _{G}(t)$. We adopt a one-for-one non-dispatchable energy buyback method for both energy trading prices, to value the energy imports and exports equally, from the CES provider and the grid by prosumers \cite{martin1}. Also, we assume there is no competition among prosumers to access the CES system.
	
	The literature lacks real data about the CES provider's energy trading price. Thus, we exploit three different energy price signals for the CES provider to compare three ETSs, and select the ETS which delivers the most equitable economic benefits for prosumers and the CES provider. The relationship between $\lambda _{G}(t)$ and $\lambda _{C}(t)$ in each ETS is given by (\ref{eq:16}a)-(\ref{eq:16}c).
	\vspace{-0.5em}
	\begin{subequations}\label{eq:16}
		\begin{empheq}[left= \lambda _{C}(t)\text{=}\empheqlbrace]{align}
			\lambda _{G,avg}=\frac{\sum_{t=1}^{24}\lambda_{G}(t)}{24} \hspace{8mm} in \hspace{2mm} ETS \hspace{2mm} 1 \hspace{18mm}\\
			\delta\lambda _{G}(t) \hspace{11mm} for \hspace{2mm} \delta > 1 \hspace{2mm} in \hspace{2mm} ETS \hspace{2mm} 2 \hspace{19.3mm} \\ 
			\delta\lambda _{G}(t) \hspace{5mm} for \hspace{2mm} 0< \delta < 1 \hspace{2mm} in \hspace{2mm} ETS \hspace{2mm} 3 \hspace{18mm}
		\end{empheq}
	\end{subequations}

	\section{{\normalsize Stochastic modeling}}
	\vspace{-0.5em}
	In this section, we model the uncertainty of energy consumption and PV generation of prosumers using the normal PDF. Also, we describe how we generate scenarios using the RWM and how the K-Means clustering method is used to reduce the number of scenarios to reduce the computational complexity of our optimization problems.
	\vspace{-0.5em}
	\subsection{Uncertainty of energy consumption and PV generation of prosumers}	
	
	We model the uncertainty of real and reactive energy consumption and PV generation using the normal distribution $f^{N}_{L, Q, PV,t}(.)$ described by (\ref{eq:17}). Its mean, standard deviation, and sample values as functions of prosumers and time are denoted by $\mu^{t}_{u}$, $\sigma^{t}_{u}$ and $X^{t}_{u}$, respectively. For real energy consumption, $\mu^{t}_{u}=\mu^{L,t}_{u}$, $\sigma^{t}_{u}=\sigma^{L,t}_{u}$ and $X^{t}_{u}=X^{L,t}_{u}$, for reactive energy consumption, $\mu^{t}_{u}=\mu^{Q,t}_{u}$, $\sigma^{t}_{u}=\sigma^{Q,t}_{u}$ and $X^{t}_{u}=X^{Q,t}_{u}$, and for PV generation, $\mu^{t}_{u}=\mu^{PV,t}_{u}$, $\sigma^{t}_{u}=\sigma^{PV,t}_{u}$ and $X^{t}_{u}=X^{PV,t}_{u}$.
	\vspace{-0.7em}
	\begin{equation}\label{eq:17}
		\begin{split}
			f^{N}_{L,Q,PV,t}(X)=\frac{1}{\sigma^{t}_{u} \sqrt{2\pi }}e^{-0.5\left ( \frac{X^{t}_{u}-\mu^{t}_{u} }{\sigma^{t}_{u} } \right )^2}  \hspace{4mm} \forall  u \in \mathcal{U}, \hspace{1mm}  t \mathcal{\in T}
		\end{split}	
	\end{equation}
	
	\subsection{Scenario-based stochastic program}
	
\begin{figure}[h!]
		\centering
		\includegraphics[width=5.5in,height=2.3in]{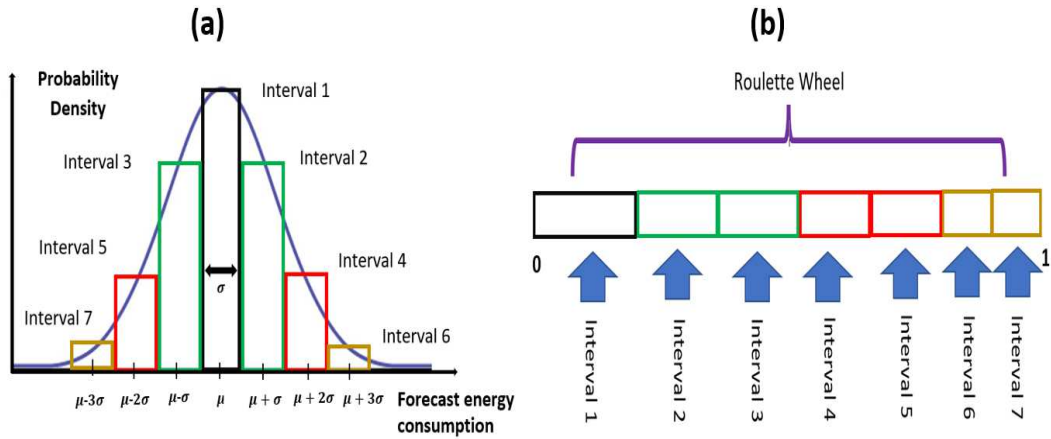}
		\caption{(a) Discretization of normal PDF into seven intervals, (b) Roulette wheel partitioned into seven intervals, each with a width of the normalized probability of the corresponding energy consumption.}
		\label{fig: Fig. 2}
		\vspace{-1em}
	\end{figure}

	Since the normal distribution is a continuous PDF, it represents an infinite number of realizations for the random variables. Here, a realization refers to a sample energy consumption or PV generation of a prosumer at a given time. A large number of realizations can model the uncertainty better at the expense of a large computational burden. Hence, we approximate the normal PDF as a discrete function by a finite number of realizations to eliminate the similar and less probable realizations and to reduce the complexity of uncertainty modeling. Thus, we use the RWM proposed in \cite{aghaei2013mip,amjady2009stochastic} to approximate the normal PDF as a discrete function having seven intervals, each with a width of a standard deviation $\sigma$, and centered around each realization. For instance, when the historical real energy consumption is $\mu$, and its standard deviation is $\sigma$, intervals 1-7 of the normal PDF are centered around the seven realizations $\mu$, $\mu + \sigma, \mu - \sigma, $ $\mu + 2\sigma,  \mu - 2\sigma, $ $\mu + 3\sigma$ and $\mu - 3\sigma$ as shown in Fig. \ref{fig: Fig. 2}(a). The same approach is used to model the uncertainty of reactive energy consumption and PV generation by discretizing normal PDFs separately for each of them. For these realizations of PV generation and energy consumption, we obtain their corresponding probabilities and generate scenarios as follows.
	
	\vspace{-0.7em}
	\begin{itemize}
		\item Do Step 1 to Step 4 
		$\forall u \in \mathcal U $, $t \in \mathcal{T}$
	\end{itemize}
	
	\vspace{-0.5em}
	Step 1: Calculate the seven realizations for each real and reactive energy consumption and PV energy generation. In this paper, we assume $\sigma=0.02\mu$. The computed realizations are taken as the midpoints of the seven $\sigma$-wide intervals in the normal distributions. In this way, the continuous normal PDFs are discretized and approximated as discrete functions. Then, the probability density for each realization is calculated by (\ref{eq:17}). Once the probability densities are available, the probability for the occurrence of each PV generation, real, and reactive energy consumption are found by taking the product of probability density and the width $\sigma$ of each interval.
	
	Step 2: Since the continuous PDFs are approximated by discretizing them, the sum of probabilities obtained after the discretization will only be close to unity but not exactly equal to one. To have the sum of the probabilities equal to unity, we normalize the calculated probabilities one by one, by taking the sum of probabilities and dividing each probability by the sum. This is done for the seven realizations obtained from each PDF separately.
	
	Step 3: We exploit the RWM explained in \cite{aghaei2013mip} to construct three roulette wheels in the range [0,1], each having seven intervals. Then, we assign the normalized seven probabilities obtained from each PDF to the [0,1] range. Hence, each interval has a width of the normalized probability of the corresponding real and reactive energy consumption and PV generation. The construction of a roulette wheel corresponds to the discretization of the normal PDF in Fig. \ref{fig: Fig. 2}(a), and a sample roulette wheel obtained like that is shown in Fig. \ref{fig: Fig. 2}(b).
	
	Step 4: Generate $N_m=\left | \mathcal{M} \right |$ number of random numbers between 0 and 1, which follow the uniform distribution, to guarantee the random numbers are generated without any bias.
	\vspace{-0.5em}
	\begin{itemize}
		\item Do Step 5 
		$\forall u \in \mathcal U $, $t \in \mathcal{T}$, $m \in \mathcal{M} $
	\end{itemize}
	
	Step 5: Assign each random number to the three roulette wheels according to their magnitudes. Select $\phi^{L,t}_{u,m},\phi^{Q,t}_{u,m}, \phi^{PV,t}_{u,m}$,  $e_{u,m}^{L}(t)$, $e_{u,m}^{Q}(t)$, and $ e_{u,m}^{PV}(t)$ from the roulette wheels corresponding to the value of the random number, where $\phi^{L,t}_{u,m}$, $\phi^{Q,t}_{u,m}$, and $\phi^{PV,t}_{u,m}$ are the normalized probabilities of $e_{u,m}^{L}(t)$, $e_{u,m}^{Q}(t)$ and $e_{u,m}^{PV}(t)$, respectively. In this way, the set of initial scenarios is obtained.
	\vspace{-0.5em}
	\begin{itemize}
		\item Do Step 6
		$\forall t \in \mathcal T $, $m \in \mathcal{M}$
	\end{itemize}
	\vspace{-0.5em}
	Step 6: Using the values found in Step 5, calculate the overall probability $\Omega_{m,t}$ in (\ref{eq:18}), which gives the probability for the occurrence of scenario $m$ at time $t$. 
	\vspace{-0.5em}
	\begin{equation}\label{eq:18}
		\Omega_{m,t}=\frac{\left( \prod_{\alpha=1}^{N} \left( \prod_{u\in U_{\alpha}}^{} \phi^{L,t}_{u,m}\phi^{Q,t}_{u,m}\phi^{PV,t}_{u,m}\right)\right)}{\sum_{m=1}^{N_m}\left( \prod_{\alpha=1}^{N} \left( \prod_{u\in U_{\alpha}}^{} \phi^{L,t}_{u,m}\phi^{Q,t}_{u,m}\phi^{PV,t}_{u,m}\right)\right)}
	\end{equation}    
	\vspace{-0.5em}
	\begin{itemize}
		\item Do Step 7  $\forall u \in \mathcal U $, $t \in \mathcal{T}$, $s \in \mathcal{S} $
	\end{itemize}
	\vspace{-0.5em}
	Step 7: A scenario reduction approach is essential in scenario-based stochastic programs to keep the problem tractable while sustaining a fair approximation for the uncertainty. Thus, the initially generated scenarios $\mathcal{M}$, are then reduced to $\mathcal{S}$ ($N_s=\left | \mathcal{S} \right |$) using the K-Means clustering algorithm mentioned in \cite{scarlatache2012using}. For this, the K-Means algorithm is applied to the sample values of real and reactive energy consumption and PV generation of prosumers separately and obtains a new set of values for $e_{u,s}^{L}(t)$, $e_{u,s}^{Q}(t)$ and $e_{u,s}^{PV}(t)$. Then, the new overall probability $\omega_{s,t}$  for the occurrence of scenario $s$ at time $t$ is computed by (\ref{eq:19}). The numerical values found for $\omega_{s,t}$, $e_{u,s}^{L}(t)$, $e_{u,s}^{Q}(t)$, and $ e_{u,s}^{PV}(t)$ $\forall u \in \mathcal{U}$, $t \mathcal{\in T}, s \mathcal{\in S}$ are then fed into the system models and optimization framework discussed in Section 2 and 4.
	\vspace{-0.5em}
	\begin{equation}\label{eq:19}
		\omega_{s,t}=\frac{\Omega_{s,t}}{\sum_{s=1}^{N_s}\Omega_{s,t}}
	\end{equation}
	\vspace{-10mm}
	\section{{\normalsize Multi-Objective Optimization Framework}}
	\vspace{-0.5em}
	In this paper, we aim to minimize the investment cost of the CES provider and the operating costs of prosumers and the CES provider to benefit both stakeholders simultaneously. This section describes the objective functions and presents a summary of the algorithm of our optimization framework.
	
	\subsection{{\normalsize Objective Functions}}
	
	\subsubsection{{\normalsize Minimizing the investment cost of the CES provider}} 
	
	The cost incurred by the CES provider for investing in the CES system can be expressed by (\ref{eq:20}), where the first term of it is the investment cost for the CES-rated power and the latter for the capacity of the CES system.
	\vspace{-0.5em}
	\begin{equation}\label{eq:20}
		f_{inv,C}= \rho^{CES}(C^{CES}_{R}p_{\alpha}^{CES,R}+C^{CES}_{C}E_{\alpha}^{CES,C})
	\end{equation}
	
	Here, $ \rho^{CES}=\frac{d(1+d)^\tau}{(1+d)^\tau-1}$, where $\rho^{CES}$, $d$ and $\tau$ are the annual cost of the CES system, discount rate and the CES system lifetime, respectively. Also, $ C^{CES}_{R}$ and $ C^{CES}_{C}$ are the CES system investment cost per kW (in $AUD/kW$) and per kWh (in $AUD/kWh$), respectively.
	\vspace{-0.5em}
	\subsubsection{{\normalsize Minimizing the expected annual operating cost of the CES provider}} The expected annual cost of operating the CES system is given by (\ref{eq:21}), and its first term is the cost of trading energy with prosumers and the second term is the cost of trading energy with the grid by the CES provider. In (\ref{eq:21}), the negative sign in the first term is due to the sign convention described in Section 2.2, and $e_{CES,s}^{G}(t) $ is the energy trade by the CES system with the grid, where $e_{CES,s}^{G}(t)=\sum_{\alpha=1}^{N}\left \{\sum_{u \in \mathcal{U}}e_{u,s}^{CES}(t) +e_{\alpha,s}^{CES,ch}(t)-e_{\alpha,s}^{CES,dis}(t)  \right \} $. Note that, $f_{op,C} <0$ implies revenue for the CES provider.
	\vspace{-0.5em}
	\begin{equation}\label{eq:21}
		f_{op,C}=\sum_{t \mathcal{ \in T}} \sum_{s=1}^{N_s}\omega_{s,t}\left \{ -\lambda _{C}(t)\sum_{\alpha=1}^{N}\sum_{u \in \mathcal{U}}e_{u,s}^{CES}(t) +\lambda _{G}(t)e_{CES,s}^{G}(t) \right \}
	\end{equation}

	\subsubsection{{\normalsize Minimizing the expected annual operating costs of prosumers}} 
	
	Prosumers incur a cost for trading energy with the CES system and the grid, which is jointly named as prosumers' operating costs, and this is given by (\ref{eq:22}). Its first term denotes the energy trading cost with the grid, and the latter is the energy trading cost with the CES system.
	\vspace{-0.5em}
	\begin{equation}\label{eq:22}
		f_{op,P}=  \sum_{t \mathcal{ \in T}} \sum_{s=1}^{N_s}\omega_{s,t}\biggl\{ \lambda _{G}(t)\sum_{\alpha=1}^{N}\sum_{u \in \mathcal{U}} e_{u,s}^{G}(t) + \lambda _{C}(t)\sum_{\alpha=1}^{N}\sum_{u \in \mathcal{U}}e_{u,s}^{CES}(t) \biggl \} 	
	\end{equation}

	\subsection{{\normalsize Optimization framework and solution method}}
	
	The optimization problems involving multi-objectives can be solved using methods, such as the weighted sum method and the hierarchical ($\epsilon$-constraint) method. In this paper, we use the $\epsilon$-constraint method as it is the recommended method for optimization problems with linear objective functions \cite{grodzevich2006normalization}. In this method, first, each objective function is ranked according to its importance. Then each objective is minimized one by one subject to a set of constraints and additional constraints formed by a prescribed fraction $\epsilon$ of the optimal value obtained for the higher-ranked objective functions. For instance, in case of a problem with three objective functions written in descending order according to their importance as $f_1$, $f_2$, and $f_3$, we first minimize $f_1$ such that $x \in \Psi $, where $\Psi $ is the feasible region. If the optimal value of $f_1$ is $f_{1}^*$, as the next step, we minimize $f_2$ such that $x \in \Psi $ and $f_1(x) \leq $ $f_{1}^*( 1 + \epsilon_{1})$. Here, $\epsilon_{1}$ is the prescribed fraction of $f_{1}^*$ which is decided by the decision maker. As the final step, we minimize $f_3$ such that $x \in \Psi $, $f_1(x) \leq $ $f_{1}^*(1 + \epsilon_{1})$ and $f_2(x) \leq $ $f_{2}^*(1 + \epsilon_{2})$, given that $f_{2}^*$ is the optimal solution of $f_{2}$ in the previous step. This method overcomes the deficiency of the weighted sum method which tends to give corner solutions of the Pareto frontier, which are not the practically desirable outcomes. Also, this technique makes sure the minimum value of the higher-ranked objective functions obtained when solved as single objective functions, will not deviate significantly from those values in the process of minimizing the less important objectives. Further explanation about the $\epsilon$-constraint method can be found in \cite{grodzevich2006normalization}.
	
	In this paper, we rank $f_{inv,C}$, $f_{op,C}$, and $f_{op,P}$ in descending order according to their importance. Let $\mathbf{x=(A_{\alpha},p_{\alpha}^{CES,R},E_{\alpha}^{CES,C}, e_{\alpha}^{CES,ch},e_{\alpha}^{CES,dis},e_{u}^{CES},e_{u}^{G})}$ be the decision variable vector with its elements being the vector of optimal CES system node, rated power, capacity, charging energy, discharging energy, prosumers' energy trade with the CES system and the grid, respectively. The feasible set is given by \textit{X}. For ETS 1, we solve (\ref{eq:23}) subject to (\ref{eq:1})-(\ref{eq:15}), (\ref{eq:16}a) and (\ref{eq:24}). Here, $f_{inv,C}^*$ is the optimal value of $f_{inv,C}$ obtained by solving (\ref{eq:20}) subject to (\ref{eq:1})-(\ref{eq:15}), (\ref{eq:16}a), and $f_{op,C}^*$ is the optimal value of $f_{op,C}$ obtained by solving (\ref{eq:21}) subject to (\ref{eq:1})-(\ref{eq:15}), (\ref{eq:16}a) and (\ref{eq:24}a). Additionally, we take $\epsilon_{1}=\epsilon_{2}=0.2$. This implies the maximum we can sacrifice is 20\% of the optimal values of $f_{inv,C}$ and $f_{op,C}$ to minimize $f_{op,P}$. Similarly, we solve (\ref{eq:23}) subject to (\ref{eq:1})-(\ref{eq:15}), (\ref{eq:16}b) and (\ref{eq:24}) for ETS 2, and (\ref{eq:23}) subject to (\ref{eq:1})-(\ref{eq:15}), (\ref{eq:16}c) and (\ref{eq:24}) for ETS 3. The optimization problem of each ETS is solved as a mixed-integer linear program (MILP). The implementation of the overall optimization framework is succinctly given in Algorithm 1.
	\vspace{-0.7em}
	\begin{equation}\label{eq:23}
		\begin{split}
			\underset{\mathbf{x} \in X}{min} \hspace{3mm} f_{op,P}
		\end{split}
	\end{equation}
	\vspace{-2em}
	\begin{subequations}\label{eq:24}
		\begin{equation}\label{eq:24a}
			\textit{s.t.} \hspace{3mm} f_{inv,C} \leq f_{inv,C}^*(1 + \epsilon_{1}),
		\end{equation}
		\begin{equation}\label{eq:24b}
			\hspace{2mm} f_{op,C} \leq f_{op,C}^*(1 + \epsilon_{2})
		\end{equation}
	\end{subequations}
	\vspace{-1.5em}
	\begin{algorithm}
		\centering
		\caption{Algorithm to Run the Multi-objective Stochastic Optimization Framework.}
		\label{algorithm:1}
		\begin{algorithmic}[1]
			\STATE Input $\mu^{L,t}_{u}, \mu^{Q,t}_{u}, \mu^{PV,t}_{u} $ $\forall u \in \mathcal{U}, t \mathcal{\in T}$ 
			\STATE Execute Steps 1 to 7 detailed in Section 3.2 to model the uncertainty of real and reactive energy consumption and PV generation of prosumers.
			\STATE Return $\omega_{s,t}$, $e_{u,s}^{L}(t)$, $e_{u,s}^{Q}(t)$, and $ e_{u,s}^{PV}(t)$ $\forall u \in \mathcal{U}, t \mathcal{\in T}, s \mathcal{\in S} $.
			\STATE Solve the described optimization problems for ETS 1-3, as MILPs.
			
		\end{algorithmic}
	\end{algorithm} 
	\vspace{-2em}	
	\section{{\normalsize Numerical and Simulation Results}}
	
\begin{figure}[h]
		\centering
		\includegraphics[width=3.0in,height=1.7in]{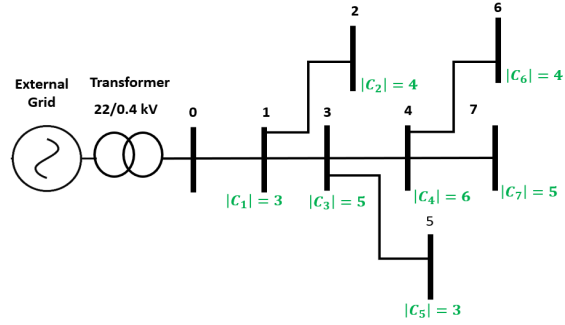}
		\caption{7-Node LV radial feeder with the number of prosumers marked at each node.}
		\label{fig: Fig. 3}
		\vspace{-1em}
	\end{figure}

	For simulations, a radial distribution network with 7 nodes given in Fig. \ref{fig: Fig. 3} was considered, and its data can be found in \cite{zeraati2016distributed}. The historical PV generation and real and reactive energy consumption data of 30 residential prosumers from an Australian community, for a period of 1 year, measured in 1-hour time intervals were used for simulations \cite{ausgridaustralia}. Also, we randomly assigned the 30 prosumers to each node except for the slack node (see Fig. \ref{fig: Fig. 3}). Therefore, $\sum_{\alpha=1}^{N}\left | U_{\alpha} \right |=30$. As model parameters, we assumed $ V_{0}=1 p.u.$, $\underline{V}=0.95 p.u.$, $\overline{V}=1.05 p.u.$, $\overline{p}^{CES}=200 kW$, $\underline{E}^{CES}=50 kWh$, $\overline{E}^{CES}=1000 kWh$, $\mu^{ch}=\mu^{dis}=0.98$, $\underline{\sigma}=0.05$, $\overline{\sigma}=1$, $\theta=0.0001 kWh$, $\Delta t=1 h$, $d=0.1$, and $\tau=12.5$ years. We took $C^{CES}_{R}=463 AUD/kW$, and $C^{CES}_{C}=795 AUD/kWh$ assuming Li-ion as the CES technology \cite{zakeri2015electrical}. Simulations for the three ETSs were implemented in Python-Pyomo and solved using the mixed-integer linear solver from CPLEX.
	\vspace{-1em}
	\subsection{{\normalsize Impact of ETS on planning and expected operating costs}}
	
\begin{figure}[b]
		\centering
		\includegraphics[width=3.0in,height=1.6in]{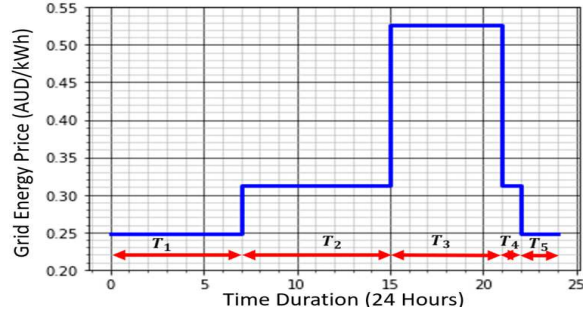}
		\caption{ToU grid energy trading price $\lambda_{G}(t)$.}
		\label{fig: Fig.4}
		\vspace{-1em}
	\end{figure}

	In this section, we analyze how the costs for prosumers and the CES provider vary based on the type of ETS, and thereby select the ETS which delivers the most equitable economic benefits for both of them. In simulations, a time-of-use (ToU) grid energy price $\lambda_G(t)$ shown in Fig. \ref{fig: Fig.4} was used \cite{origin}. In ETS 1, $\lambda_{C}(t)=\lambda_{G,avg}=0.34180 \hspace{1mm} AUD/kWh $, in ETS 2 we take $\delta=1.5$ since $\lambda_{C}(t) > \lambda_{G}(t) $, and in ETS 3 we use $\delta=0.5$ as $\lambda_{C}(t) < \lambda_{G}(t) $. We considered 50 initial scenarios (i.e. $N_m=50$), which were then reduced to 10 (i.e. $N_s=10$) by using the K-Means clustering algorithm for each ETS. Table  \ref{table:1} gives a summary of the numerical results obtained for each ETS. 
	
	\begin{table}[t]
		\centering
		\caption{Planning and operational results for different ETSs (With $N_s=10)$.}
		\label{table:1}
		\tiny
		\renewcommand{\arraystretch}{2.0}
		\begin{tabular}{|c|c|c|c|c|c|c|}
			\hline
			& \begin{tabular}[c]{@{}c@{}}Optimal \\ CES \\ node\end{tabular} & \begin{tabular}[c]{@{}c@{}}Optimal CES\\ capacity (kWh)\end{tabular} & \begin{tabular}[c]{@{}c@{}}Optimal CES\\ rated power\\ (kW)\end{tabular} & \begin{tabular}[c]{@{}c@{}}CES provider's\\ investment\\ cost (AUD)\end{tabular} & \begin{tabular}[c]{@{}c@{}}CES provider's\\ expected annual \\ operating revenue \\ (AUD)\end{tabular} & \begin{tabular}[c]{@{}c@{}}Prosumers'\\ expected annual \\ operating costs \\(AUD)\end{tabular} \\ \hline
			No CES & N/A                                                         & N/A                                                                  & N/A                                                                      & N/A                                                                              & N/A                                                                               & 101146                                                                       \\ \hline
			ETS 1  & 7                                                           & 414                                                                  & 156                                                                      & 57702                                                                            & 38024 (41\%)$^1$                                                                  & 45230 (14\%)$^1$                                                              \\ \hline
			ETS 2  & 7                                                           & 414                                                                  & 156                                                                      & 57702                                                                            &  64676                                                                             & 
			52870
			\\ \hline
			ETS 3  & 7                                                           & 414                                                                  & 156                                                                      & 57702                                                                            & 34089 (47\%)$^1$                                                                   &   32914 (38\%)$^1$                                                                 \\ \hline
			
			\multicolumn{7}{l}{$^1$ Percentage values were computed with respect to the maximum operating revenue or costs.}
		\end{tabular}
	\end{table}
	
	The optimal planning aspects of the CES system, namely its location, capacity, and rated power are the same for the three ETSs. This happens because investment decisions are prevalent over the operating decisions in our case study for three ETSs. Hence, the planning costs are the same for the three ETSs.
	
	The operational objectives are to minimize the operating cost (i.e., maximize the operating revenue) of the CES provider, and minimize the operating costs of prosumers. Table  \ref{table:1} shows that the revenue of the CES provider is the highest in ETS 2 and the lowest in ETS 3. In ETS 2, as $\lambda_{CES}(t) >$ $\lambda_{G}(t)$ $\forall t \mathcal{\in T}$, the CES provider sells its energy to the grid and prosumers at a higher price, and imports energy from the grid at a lower price resulting in the maximum revenue for the CES provider, whereas in ETS 3, since $\lambda_{CES}(t) <$ $\lambda_{G}(t)$ $\forall t \mathcal{\in T}$, the CES provider has to import high priced energy from the grid, and earn less from its exports to prosumers and the grid. In contrast, prosumers' operating costs are the highest in ETS 2 and the lowest in ETS 3. Hence, ETS 2 generates the highest operating revenue for the CES provider and the highest operating costs for prosumers, while ETS 3 is the opposite. Also, prosumers' operating costs are reduced in each ETS after integrating the CES system. Nevertheless, in ETS 1, the operating costs/revenue for both beneficiaries are moderate compared to their values in ETS 2 and 3. Hence, ETS 1 spreads the economic benefits between the CES provider and prosumers more equitably than the other two ETSs.
	
	\vspace{-0.5em}
	\subsection{{\normalsize Comparison of the economic benefits of prosumers and the CES provider}}
	
	In this section, we discuss how the economic benefits for prosumers and the CES provider vary with the optimal CES system planning and operation in each ETS. Also, we analyze how energy transactions between prosumers, the grid, and the CES system occur in the three ETSs. For this, we do the analysis for a 
	selected day with a duration of 24 hours.
	
	\subsubsection{{\normalsize ETS 1: When $\lambda_{CES}(t)$=$\lambda_{G,avg}, \hspace{2mm} \forall t \mathcal{\in T}$}} 
\begin{figure}[h!]
		\centering
		\includegraphics[width=4.5in,height=5.6 cm]{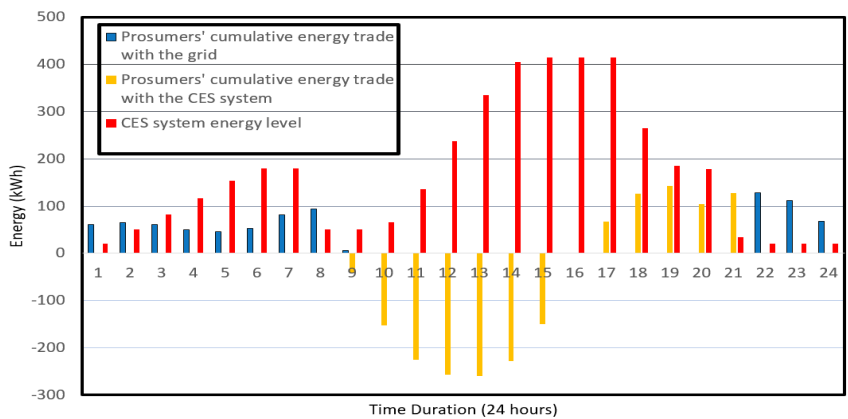}
		\caption{Energy trade by prosumers with the grid and the CES system, and the temporal variation of the CES system energy in ETS 1 (for $N_s=10$ and $\lambda_{CES}(t)=\lambda_{G,avg}$). }
		\label{fig: Fig. 5}
		\vspace{-1em}
	\end{figure}

Fig. \ref{fig: Fig. 5} depicts how prosumers trade energy with the grid (blue plot) and the CES system (orange plot), and the CES system energy level variation with time (red plot). In Fig. \ref{fig: Fig. 5}, the energy imports and exports by prosumers are seen as positive and negative values, respectively. During $T_{3}$, $\lambda_{CES}(t) < \lambda_{G}(t) $, and during $T_{1},T_{2}$ $T_{4},T_{5}$,  $\lambda_{CES}(t) > \lambda_{G}(t) $. According to Fig. \ref{fig: Fig. 5}, prosumers import energy only from the grid during $T_{1}$. Since $\lambda_{CES}(t) > \lambda_{G}(t)$ during this time period, it is not economically beneficial for prosumers to import expensive energy from the CES system. During $T_2$, in which the time period with high PV generation, prosumers export the excess PV energy to the CES system, as $\lambda_{CES}(t) > \lambda_{G}(t) $, resulting in higher revenue from the CES provider. Since $\lambda_{CES}(t) < \lambda_{G}(t)$ during $T_3$, prosumers import energy from the CES system, so they have to pay less for the imported energy. During $T_4$ and $T_5$, prosumers import energy only from the grid as $\lambda_{CES}(t) > \lambda_{G}(t)$. Considering the temporal variation of the CES system energy level, it is observed that the CES system charges during $T_1$ from the low-priced grid energy, and it discharges partially by the end of $T_1$. The CES system continues to charge during $T_2$ until it reaches its full capacity using the excess PV generation of prosumers. This is evident as the CES system energy level has reached its full capacity of 414 kWh during $T_2$. During $T_3$, the CES system exports its energy back to prosumers, and at the end of the day, the CES system reaches its initial energy level.
	
	\subsubsection{{\normalsize ETS 2: When $\lambda_{CES}(t)$=$1.5\lambda_{G}(t), \hspace{2mm} \forall  t \mathcal{\in T}$}} 
	
\begin{figure}[h!]
		\centering
		\includegraphics[width=4.5in,height=5.5 cm]{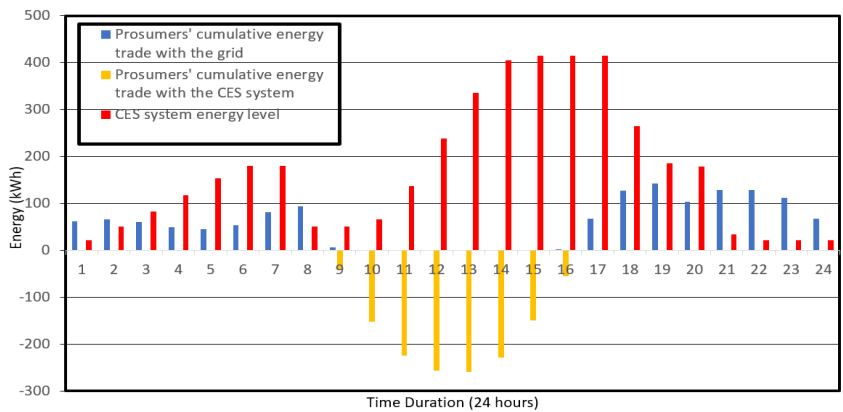}
		\caption{Energy trade by prosumers with the grid and the CES system, and the temporal variation of the CES system energy in ETS 2 (for $N_s=10$ and $\lambda_{CES}(t)=1.5\lambda_{G}(t)$).}
		\label{fig: Fig. 6}
		\vspace{-1em}
	\end{figure}

	In ETS 2, as illustrated in Fig. \ref{fig: Fig. 6}, prosumers import energy only from the grid during  $T_{1}, T_{3}$ $T_{4}$, and $T_{5}$, the times of the day in which the PV generation is insufficient to fulfill the energy demand of prosumers and during $T_{2}$, prosumers export their excess PV generation to the CES system. Since $\lambda_{CES}(t)$ $>\lambda_{G}(t)$, importing energy from the grid gives lesser costs, and exporting energy to the CES system generates high earnings for prosumers. In this ETS also, the CES system charges from the grid energy and partially discharges by the end of  $T_1$, and again charges from the excess PV energy of prosumers during  $T_2$, and continues to discharge during  $T_3$, $T_4$ and $T_5$ until it reaches its initial energy level at the start of the day. 
	\vspace{-2mm}
	\vspace{-2mm}
	\subsubsection{{\normalsize ETS 3: When $\lambda_{CES}(t)$=$0.5\lambda_{G}(t), \hspace{2mm} \forall  t \mathcal{\in T}$}} 
	
\begin{figure}[h!]
		\centering
		\includegraphics[width=4.5in,height=5.6 cm]{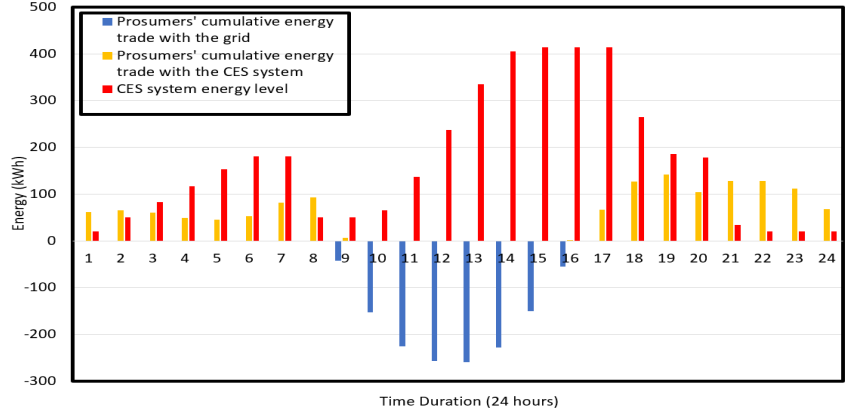}
		\caption{Energy trade by prosumers with the grid and the CES system, and the temporal variation of the CES system energy in ETS 3 (for $N_s=10$ and $\lambda_{CES}(t)=0.5\lambda_{G}(t)$).}
		\label{fig: Fig. 7}
		\vspace{-1em}
	\end{figure}

	In contrast to ETS 2, prosumers import energy only from the CES system during  $T_{1}, T_{3}$ $T_{4}, T_{5}$, and during $T_{2}$, prosumers export their excess PV generation to the grid. This is observed in Fig. \ref{fig: Fig. 7}. Since $\lambda_{CES}(t)$ $ <\lambda_{G}(t)$, importing energy from the CES system generates lesser costs, and exporting energy to the grid provides high earnings for prosumers. The temporal variation of the CES system energy level is also shown in Fig. \ref{fig: Fig. 7}.
	\vspace{-4mm}
	\vspace{-2mm}
	\subsection{{\normalsize Evaluation of the quality of the stochastic solution}}
	
	To assess the accuracy of the solutions obtained from our stochastic programs, we use the out-of-sample simulation method discussed in \cite{Roald2023power}. For simplicity, we consider only the results obtained in ETS 1. Then we generate another 20 scenarios and form a new scenario set $\mathcal{D}$ such that $\mathcal{M}\cap \mathcal{D}=\O$, $\mathcal{S}\cap \mathcal{D}=\O$, and $\left | \mathcal{S} \right | < \left | \mathcal{D} \right |$. In the out-of-sample simulation method, we fix the values of the planning decision variables in $f_{inv,C}$ and deterministically solve for $f_{op,C}$ and $f_{op,P}$ for 20 times excluding the scenario dependency of (\ref{eq:1})-(\ref{eq:15}), (\ref{eq:21}) and (\ref{eq:22}), and calculate the average of the operational objectives. The average operating revenue of the CES provider and the operating costs of prosumers were found as AUD 40079 and AUD 48621, respectively. To measure the quality of the stochastic solution, we use the "Value of Stochastic Solution (VSS)", which gives the difference between the deterministic and the stochastic solution of the objective function. In ETS 1, VSS=$6.97\%$, which implies a high-quality stochastic solution since the value is small.
	\vspace{-1.0em}
	\subsection{{\normalsize Pareto frontier}}
	
\begin{figure}[h]
		\centering
		\includegraphics[width=5in,height=6.3 cm]{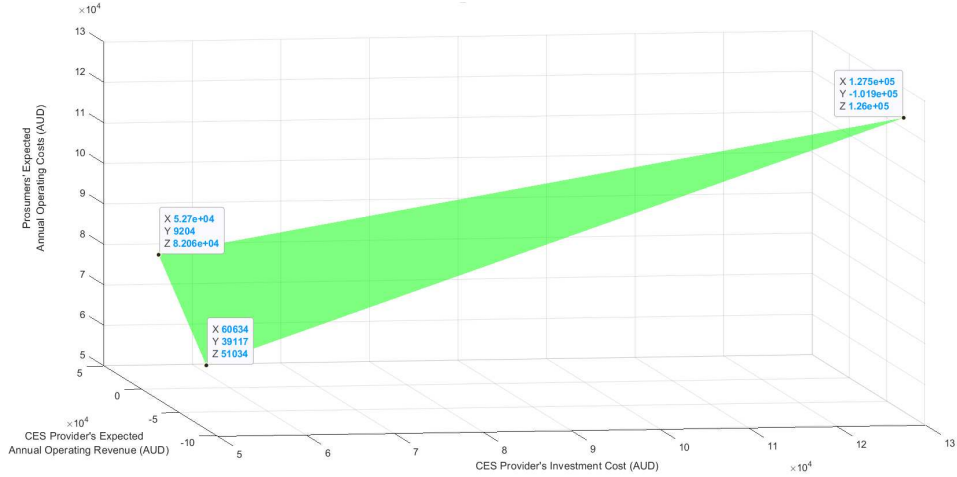}
		\caption{Pareto frontier of the multi-objective stochastic optimization model in ETS 1.}
		\label{fig: Fig. 8}
		\vspace{-1em}
	\end{figure}

	In this paper, we optimize the CES system planning and its operation to minimize prosumers' operating costs while minimizing the investment and operating costs of the CES provider. Hence, our work considers multi-objective optimization, which generates a set of optimal solutions instead of a single unique solution. The set of solutions obtained forms the Pareto frontier. Since we consider three objectives, it forms a Pareto frontier plane and the one obtained in ETS 1 is shown in Fig. \ref{fig: Fig. 8}. The corner points of the Pareto frontier plane are the utopia (i.e. lower bounds of the Pareto frontier) and the nadir (i.e. upper bounds of the Pareto frontier) values, which are marked in Fig. \ref{fig: Fig. 8}. Since we exploited the  $\epsilon$-constraint method to find the optimal solution of our multi-objective optimization framework, it gives a solution point on the Pareto frontier avoiding the corner points as the solution. It is essential to avoid the corner points of the Pareto frontier as the optimal solution, since they are obtained by degrading at least one of the objectives significantly in the quest to minimize the other objectives \cite{grodzevich2006normalization}. The obtained solution in ETS 1 for the CES provider's investment cost and operating revenue and prosumers' operating costs are AUD 57702, AUD 38024, and AUD 45230, which are the coordinates of a point on the Pareto frontier plane, that balances benefits between prosumers and the CES provider.
	
	\vspace{-1em}
	\section{{\normalsize Conclusion \& Future Work }}
	\vspace{-2mm}
	In this paper, we have evaluated how the planning of community energy storage (CES) systems under uncertainty impacts the economic benefits for prosumers and the CES provider. For this, we have proposed a multi-objective stochastic optimization framework to select the CES solution (optimal CES system planning and the ETS)  that delivers the most economically equitable outcome to the CES provider and prosumers. We model uncertainties of PV generation, and real and reactive energy consumption using stochastic scenarios generated initially by a roulette wheel mechanism (RWM) and then aggregated by the K-Means clustering algorithm to reduce the computational complexity of our optimization problems. 
	
	Our experiments show that in the ETS where the CES provider's energy trading price is higher than the grid energy trading price, the CES provider is benefited more, and in the ETS where the CES provider's energy trading price is lower than the grid energy trading price, prosumers are benefited more. In the ETS where the average value of the grid energy trading price was used as the CES provider's energy trading price, both prosumers and the CES provider benefited more equitably than the other two ETSs.
	
	Future work includes incorporating the uncertainty of distributed energy resources allocation, competition among the CES providers to install and operate a CES system, and peer-to-peer energy trading by prosumers and multiple CESs.
	\vspace{-4mm}
	\bibliographystyle{Elsevier}

\end{document}